\documentclass{elsart}
\usepackage{epsfig,latexsym,amsfonts,amsmath,amssymb}
\def\3{\ss}

\begin{document}
\begin{frontmatter}
 
\title{
\vspace{-2.5cm}
\hfill {\normalsize \mdseries DESY 03-205} \\[-0.2cm]
{\normalsize \mdseries LTH 614} 
\hfill {\normalsize \mdseries Edinburgh 2003/24} \\[-0.2cm]
{\normalsize \mdseries HU-EP-03/86} 
\hfill {\normalsize \mdseries LU-ITP 2003/032} \\[0.4cm] 
The nucleon mass in $N_f = 2$ lattice QCD: finite size effects 
from chiral perturbation theory}

\vspace{-0.9cm}

\author[HUB]{A. Ali Khan},
\author[Dubna]{T. Bakeyev},
\author[Leipzig,Regensbg]{M. G\"ockeler},
\author[TUM]{T.R. Hemmert},
\author[Edinburgh]{R. Horsley},
\author[Liverpool]{A.C. Irving},
\author[Edinburgh]{B. Jo\'o},
\author[Zeuthen]{D. Pleiter},
\author[Liverpool]{P.E.L. Rakow},
\author[Zeuthen,Hamburg]{G. Schierholz},
\author[ZIB]{H. St\"uben}

\collab{QCDSF-UKQCD Collaboration}

\address[HUB]{Institut f\"ur Physik, Humboldt Universit\"at zu Berlin,
              D-10115 Berlin}
\address[Dubna]{Joint Institute for Nuclear Research, 141980 Dubna, Russia}
\address[Leipzig]{Institut f\"ur Theoretische Physik,
         Universit\"at Leipzig, D-04109 Leipzig, Germany}
\address[Regensbg]{Institut f\"ur Theoretische Physik,
         Universit\"at Regensburg, D-93040 Regensburg, Germany}
\address[TUM]{Physik-Department, Theoretische Physik,
          Technische Universit\"at M\"unchen, D-85747 Garching, Germany}
\address[Edinburgh]{School of Physics, University of Edinburgh, 
          Edinburgh EH9 3JZ, UK}
\address[Liverpool]{Theoretical Physics Division, 
          Department of Mathematical Sciences, \\ 
          University of Liverpool, Liverpool L69 3BX, UK}
\address[Zeuthen]{John von Neumann-Institut f\"ur Computing NIC, \\ 
         Deutsches Elektronen-Synchrotron DESY, D-15738 Zeuthen, Germany}
\address[Hamburg]{Deutsches Elektronen-Synchrotron DESY, 
                  D-22603 Hamburg, Germany}
\address[ZIB]{Konrad-Zuse-Zentrum f\"ur Informationstechnik Berlin,
              D-14195 Berlin, Germany}

\date{}

\begin{abstract}
In the framework of relativistic SU(2)$_\mathrm f$ baryon chiral 
perturbation theory we calculate the volume dependence of the nucleon mass
up to and including $O(p^4)$. Since the parameters in the resulting 
finite size formulae are fixed from the pion mass dependence of the 
large volume nucleon masses and from phenomenology, we obtain a 
parameter-free prediction of the finite size effects. We present
mass data from the recent $N_f=2$ simulations of the UKQCD and 
QCDSF collaborations and compare these data as well as published mass 
values from the dynamical simulations of the CP-PACS and JLQCD 
collaborations with the theoretical expectations. 
Remarkable agreement between the lattice data and the predictions of
chiral perturbation theory in a finite volume is found.
\end{abstract}

\end{frontmatter}
 
\section{Introduction}
\label{sect:intro}

The computation of hadron masses is one of the basic goals of lattice QCD.
However, as in the case of other observables, this computation suffers
from a number of
systematic uncertainties (not including the statistical errors): Lattice
spacing and volume are finite, and the quark masses
in the simulations are larger than in reality, in the extreme case of
the sea quarks in the quenched approximation even infinite. 
Hence the Monte Carlo results must be supplemented by several
extrapolations: the continuum extrapolation $a \to 0$, the extrapolation
to the thermodynamic limit $L \to \infty$, and the chiral extrapolation
sending the quark masses to their physical values. 

It is therefore an important issue to
study hadron masses as functions of the quark masses or, in order to
avoid problems connected with the evaluation of quark masses on the 
lattice, as functions of the pseudoscalar (``pion'') mass. 
A suitable tool for investigating the quark-mass dependence of 
physical observables like hadron masses is chiral perturbation 
theory (or, more generally, chiral effective field theory). Once
one has convinced oneself that chiral perturbation theory is applicable
for the masses used in the simulations, one can extract low-energy
constants of chiral effective field theory and extrapolate reliably
towards the physical masses. Indeed, as has recently been demonstrated,
relativistic baryon chiral perturbation theory leads to a good 
chiral extrapolation function for the nucleon mass~\cite{nmass} 
(see also Ref.~\cite{meissner}) 
connecting available lattice results with the physical value.
On the other hand, it is well known that finite size effects can 
be calculated from the same effective field theories that describe the
quark-mass dependence~\cite{fse,haleu}: As long as the volume is not
too small, the finite size effects originate from pions which ``propagate
around the spatial box''. This approach leads to the
so-called $p$ expansion, valid for small pion masses $m_\pi = O(p)$
and large spatial volumes $L^3$ with $L^{-1} = O(p)$ such that 
$m_\pi L = O(p^0)$. In this paper we work out the finite size
corrections for the nucleon mass in this framework and present 
pion and nucleon masses obtained by the UKQCD and QCDSF collaborations
in simulations with $N_f=2$ dynamical fermions. These and other 
recent nucleon mass data are compared with the formulae from chiral
perturbation theory. Preliminary results of our investigation have 
been presented in Ref.~\cite{arifa}. For similar studies of the pion 
mass and pseudoscalar decay constants see Refs.~\cite{duerr,damir}.
Results from a different, but related, chiral analysis have been given
in Ref.~\cite{adelaide}.

\section{Finite size effects at $O(p^3)$}

\begin{figure}[htbp]
\begin{center}
\epsfig{file=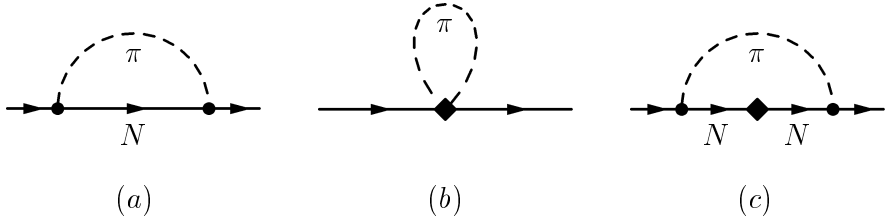,width=12cm}
\caption{One-loop graphs of NLO (a) and NNLO (b, c) contributing to 
         the nucleon mass shift. The solid circle denotes a vertex from 
	 the leading order Lagrangian, the diamond a vertex from the
	 $O(p^2)$ Lagrangian.}
\label{fig:diags}
\end{center}
\end{figure}

We follow Ref.~\cite{nmass} and employ relativistic SU(2)$_\mathrm f$
baryon chiral perturbation theory as described in Ref.~\cite{beleu}.
In this field theory the ``elementary'' degrees of freedom are the pion
and nucleon fields. For the Lagrangian and further details see 
Refs.~\cite{gasser,nmass,beleu}. 
In particular, we shall use the so-called infrared 
regularisation~\cite{beleu}, a variant of dimensional regularisation.
The leading order contribution to the shift of
the nucleon mass away from its value in the chiral limit comes from
the piece in the $O(p^2)$ Lagrangian that breaks chiral symmetry
explicitly. The next-to-leading order (NLO) contribution, i.e.\ the
$O(p^3)$ contribution, is generated by the one-loop graph (a) of 
Fig.~\ref{fig:diags}. 

The leading one-loop formula for the nucleon mass reads (in the infinite
volume)
\begin{eqnarray}
m_N & = & m_0 -4 c_1 m_\pi^2+ \left[ e_1^r(\lambda) 
 + \frac{3 g_A^2}{64 \pi^2 f_\pi^2 m_0}
   \left( 1 - 2 \ln{\frac{m_\pi}{\lambda}} \right) \right]m_\pi^4 
\nonumber \\ & {} & {}
 - \frac{3 g_A^2}{16 \pi^2 f_\pi^2}\, m_\pi^3\, 
    \sqrt{1-\frac{m_\pi^2}{4 m_0^2}} \left[\frac{\pi}{2} + 
       \arctan{\frac{m_\pi^2}{\sqrt{4m_0^2 m_\pi^2- m_\pi^4}}}\right] \,.
\label{mnp3}
\end{eqnarray}
Here and in the following, the constants $g_A$, $f_\pi$, \ldots  are to 
be taken in the chiral limit, $m_0$ denotes the nucleon mass in the 
chiral limit, and the quark mass has been replaced by the pion mass $m_\pi$ 
using the Gell-Mann-Oakes-Renner relation. The pion decay constant $f_\pi$
is normalised such that its physical value is $92.4 \, \mbox{MeV}$.
The counterterm $e_1^r(\lambda)$ is taken at the renormalisation 
scale $\lambda$, which makes the result (\ref{mnp3}) scale independent.

Expanding in powers of $m_\pi$ (up to logarithms) we find
\begin{eqnarray}
m_N & = & m_0 - 4 c_1 m_\pi^2-\frac{3 g_A^2}{32 \pi f_\pi^2} m_\pi^3 
 + \left[e_1^r(\lambda)-\frac{3 g_A^2}{64 \pi^2 f_\pi^2 m_0}
  \left( 1 + 2 \ln{\frac{m_\pi}{\lambda}} \right) \right] m_\pi^4
\nonumber \\ & {} & {}
 + \frac{3 g_A^2}{256 \pi f_\pi^2 m_0^2}m_\pi^5 + O(m_\pi^6) \,.
\label{mnp3ex}
\end{eqnarray}
In Ref.~\cite{nmass} it was shown that this expansion is a good 
approximation of the full expression (\ref{mnp3}). Note, however,
that all the terms $\propto m_\pi^3$, $m_\pi^4$, \ldots in 
Eq.~(\ref{mnp3ex}) are of the same chiral order.

At $O(p^3)$ the volume dependence of the nucleon mass comes from 
graph (a) in Fig.~\ref{fig:diags}, where the pion couples to the nucleon
through the pseudovector derivative coupling of the leading order 
Lagrangian~\cite{gasser}.
The corresponding contribution 
to the nucleon mass reads in Minkowski space
\begin{equation}
m_a = - \mathrm i \frac{3 g_A^2 m_0 m_\pi^2}{2 f_\pi^2} 
 \int_0^\infty \! \mathrm d x \int \! \frac{\mathrm d^4 p}{(2 \pi)^4} 
   \left[ p^2 - m_0^2 x^2 - m_\pi^2 (1-x) + \mathrm i \epsilon \right]^{-2} \,.
\end{equation}
Note that the rules of infrared regularisation have led us to take the
integral over the Feynman parameter $x$ from 0 to $\infty$ rather than
from 0 to 1. 
After a Wick rotation this becomes in Euclidean notation
\begin{equation}
m_a = D \int_0^\infty \! \mathrm d x \int \! \frac{\mathrm d^4 p}{(2 \pi)^4} 
   \left[ p^2 + m_0^2 x^2 + m_\pi^2 (1-x) \right]^{-2} 
\end{equation}
with
\begin{equation}
 D =  \frac{3 g_A^2 m_0 m_\pi^2}{2 f_\pi^2} \,.
\end{equation}

In a finite spatial volume of linear size $L$ (the time direction is 
left infinite) the integral over the spatial components of the loop 
momentum $\vec{p}$ is replaced by a sum over the discrete set of 
momenta allowed by the boundary conditions. Since we use periodic 
boundary conditions, the allowed momenta are of the form 
$\vec{p} = (2 \pi/L) \vec{\ell}$, where $\vec{\ell}$ is a vector with integer
components. Hence we get for the difference between the nucleon mass 
in a volume of size $L^3$ and the infinite volume nucleon mass at $O(p^3)$
\begin{eqnarray}
 & & m_N (L) - m_N (\infty) = \Delta_a(L) = \nonumber \\
 & & {} = 
 D \int_0^\infty \! \mathrm d x \int \! \frac{\mathrm d p_4}{2 \pi} 
 \left[ \frac{1}{L^3} \sum_{\vec{p}} 
  \left( \vec{p}\,^2 + p_4^2 + m_0^2 x^2 + m_\pi^2 (1-x) \right)^{-2}
\right. \nonumber \\  
 & & \quad {}  - \int \! \frac{\mathrm d^3 p}{(2 \pi)^3} \left. 
  \left( \vec{p}\,^2 + p_4^2 + m_0^2 x^2 + m_\pi^2 (1-x) \right)^{-2}
 \right] \,.
\end{eqnarray}
According to Ref.~\cite{haleu} the difference between the sum and the 
integral is finite and given by
\begin{eqnarray}
 & &  
\frac{\Gamma (r)}{L^3} \sum_{\vec{p}} \left( \vec{p}\,^2 + M^2 \right)^{-r}
- \frac{\Gamma (r)}{(2 \pi)^3} \int \! \mathrm d^3 p 
 \left( \vec{p}\,^2 + M^2 \right)^{-r} 
\nonumber \\ & & \quad {}
 = \sum_{\vec{n}} {}^\prime \, (4 \pi)^{-3/2} 
  \int_0^\infty \! \mathrm d \lambda \, \lambda^{r-5/2} 
    \exp \left( - \lambda M^2 - L^2 \vec{n}\,^2/(4 \lambda) \right) \,,
\end{eqnarray}
where the sum extends over all vectors $\vec{n}$ with integer components
excluding $\vec{n} = \vec{0}$. Here $n_i$ can be interpreted as the 
number of times the pion crosses the ``boundary'' of the lattice in the 
$i$ direction. 
With the help of 
\begin{equation}
 \int_0^\infty \! \mathrm d \lambda \, \lambda^a \, 
  \mathrm e ^{- \lambda M^2 - b/\lambda} = 
  2 \left( \frac{b}{M^2} \right)^{(1+a)/2} K_{1+a}(2 \sqrt{b M^2})
\end{equation}
(see formula 3.471.9 in~\cite{grad}) we obtain
\begin{eqnarray}
 & &  
\frac{\Gamma (r)}{L^3} \sum_{\vec{p}} \left( \vec{p}\,^2 + M^2 \right)^{-r}
- \frac{\Gamma (r)}{(2 \pi)^3} \int \! \mathrm d^3 p 
 \left( \vec{p}\,^2 + M^2 \right)^{-r} 
\nonumber \\ & & \quad {}
 = \frac{1}{4 \pi^{3/2}} \sum_{\vec{n}} {}^\prime \, 
  \left( \frac{L^2 \vec{n}\,^2}{4 M^2} \right)^{r/2 - 3/4} 
  K_{r-3/2} (\sqrt{L^2 \vec{n}\,^2 M^2}) \,,
\label{basic}
\end{eqnarray}
where $K_\nu (x)$ is a modified Bessel function.
For the numerical evaluation at smaller masses it is more advantageous
to use the relation
\begin{eqnarray}
 & &  
\frac{\Gamma (r)}{L^3} \sum_{\vec{p}} \left( \vec{p}\,^2 + M^2 \right)^{-r}
- \frac{\Gamma (r)}{(2 \pi)^3} \int \! \mathrm d^3 p 
 \left( \vec{p}\,^2 + M^2 \right)^{-r} 
\nonumber \\ & & \quad {}
 = \frac{L^{2r-3}}{(4 \pi)^r} \int_0^\infty \! \mathrm d t \,
t^{r-5/2} \exp \left( - \frac{M^2 L^2}{4 \pi} t \right)
\left[ S(1/t)^3 -1 \right]
\end{eqnarray}
with the theta function
\begin{equation}
S(x) = \sum_{n=-\infty}^\infty \mathrm e ^{- \pi n^2 x} =
\frac{1}{\sqrt{x}} S(1/x) \,.
\end{equation}

For $r=2$ we have from (\ref{basic})
\begin{eqnarray}
 & &  
\frac{1}{L^3} \sum_{\vec{p}} \left( \vec{p}\,^2 + M^2 \right)^{-2}
- \frac{1}{(2 \pi)^3} \int \! \mathrm d^3 p 
 \left( \vec{p}\,^2 + M^2 \right)^{-2} 
\nonumber \\ & & \quad {}
 = \frac{1}{4 \pi^{3/2}} \sum_{\vec{n}} {}^\prime \, 
  \left( \frac{L^2 \vec{n}\,^2}{4 M^2} \right)^{1/4} 
  K_{1/2} (\sqrt{L^2 \vec{n}\,^2 M^2}) \,.
\end{eqnarray}
Hence we can write
\begin{eqnarray}
 & & \Delta_a (L) \nonumber \\
 & & {} = 
 D \int_0^\infty \! \mathrm d x \int \! \frac{\mathrm d p_4}{2 \pi} \cdot 
 \frac{1}{4 \pi^{3/2}} \sum_{\vec{n}} {}^\prime \,
  \left( \frac{L^2 \vec{n}\,^2}{4} \right)^{1/4}
  \left( p_4^2 + m_0^2 x^2 + m_\pi^2 (1-x) \right)^{-1/4}
 \nonumber \\  
 & & \quad {}  \times   K_{1/2} \left( L |\vec{n}| 
            \sqrt{p_4^2 + m_0^2 x^2 + m_\pi^2 (1-x)} \right) \,.
\end{eqnarray}
Using
\begin{equation}
 \int_{- \infty}^\infty \! \mathrm d x \, (x^2 + a^2)^{-1/4} 
   K_{1/2} (b \sqrt{x^2 + a^2}) = \sqrt{\frac{2 \pi}{b}} K_0 (ab)
\end{equation}
(see formula 6.596.3 in~\cite{grad}) we obtain finally
\begin{eqnarray}
 & & m_N (L) - m_N (\infty) = \Delta_a(L) 
\nonumber \\
 & & {} = 
 \frac{3 g_A^2 m_0 m_\pi^2}{16 \pi^2 f_\pi^2}
 \int_0^\infty \! \mathrm d x \,
 \sum_{\vec{n}} {}^\prime \, 
  K_0 \left( L |\vec{n}| \sqrt{m_0^2 x^2 + m_\pi^2 (1-x)} \right) \,.
\label{Deltaa}
\end{eqnarray}
This is the complete $O(p^3)$ result for the volume dependence of the
nucleon mass in relativistic baryon chiral perturbation theory. Note that
the finite volume has not introduced any new parameter.

\section{Finite size effects at $O(p^4)$}

At $O(p^4)$, the nucleon self-energy receives additional contributions
from graphs (b) and (c) in Fig.~\ref{fig:diags}. Up to higher-order
corrections, the contribution to the nucleon mass from graph (c) is
cancelled by the contribution arising from the insertion of
$m_N = m_0 - 4 c_1 m_\pi^2$ in the $O(p^3)$ piece corresponding to
graph (a). Alternatively, one could replace the mass $m_0$ in the free
Lagrangian by $ m_0 - 4 c_1 m_\pi^2$ and omit graph (c)~\cite{beleu}.
Thus only graph (b) is relevant for us. The resulting formula for 
the nucleon mass in the infinite volume at $O(p^4)$ involves two new 
coupling constants $c_2$ and $c_3$ and reads~\cite{nmass,beleu}
\begin{eqnarray}
 m_N & = & m_0 - 4 c_1 m_\pi^2-\frac{3 g_A^2}{32 \pi f_\pi^2} m_\pi^3 
 + \left[e_1^r(\lambda)-\frac{3}{64 \pi^2 f_\pi^2}
    \left( \frac{g_A^2}{m_0} - \frac{c_2}{2} \right) \right.
 \nonumber \\ & {} & {} \qquad \qquad \qquad \left.
  - \frac{3}{32 \pi^2 f_\pi^2}
       \left( \frac{g_A^2}{m_0} - 8c_1 + c_2 + 4 c_3 \right)
   \ln{\frac{m_\pi}{\lambda}} \right] m_\pi^4
\nonumber \\ & {} & {}
 + \frac{3 g_A^2}{256 \pi f_\pi^2 m_0^2}m_\pi^5 + O(m_\pi^6) \,.
\label{mnp4}
\end{eqnarray}
The contribution to the nucleon mass from graph (b) 
in Fig.~\ref{fig:diags} is given by
\begin{equation} 
m_b = - \mathrm i \frac{3}{f_\pi^2} \int \! \frac{\mathrm d^4 p}{(2 \pi)^4} 
\frac{2 c_1 m_\pi^2 - (c_2 + c_3) (p^0)^2 + c_3 \vec{p}^{\,2}}
     { m_\pi^2 - p^2 - \mathrm i \epsilon}
\end{equation}
in Minkowski space. 
After a Wick rotation this becomes in Euclidean notation
\begin{equation} 
m_b = \frac{3}{f_\pi^2} \int \! \frac{\mathrm d^4 p}{(2 \pi)^4} 
\frac{2 c_1 m_\pi^2 + (c_2 + c_3) p_4^2 + c_3 \vec{p}^{\,2}}
     { m_\pi^2 + p^2} \,.
\end{equation}
Hence we get an additional contribution to the difference between the 
nucleon mass in a volume of size $L^3$ and the infinite volume nucleon 
mass:
\begin{eqnarray}
 & & \Delta_b (L) = 
  \frac{3}{f_\pi^2} \int \! \frac{\mathrm d p_4}{2 \pi} 
 \left[ \frac{1}{L^3} \sum_{\vec{p}} 
  \frac{2 c_1 m_\pi^2 + (c_2 + c_3) p_4^2 + c_3 \vec{p}^{\,2}}
     { m_\pi^2 + p_4^2 + \vec{p}^{\,2}} 
\right. \nonumber \\  
 & & \qquad \qquad \qquad \qquad {}  
  - \int \! \frac{\mathrm d^3 p}{(2 \pi)^3} \left. 
   \frac{2 c_1 m_\pi^2 + (c_2 + c_3) p_4^2 + c_3 \vec{p}^{\,2}}
     { m_\pi^2 + p_4^2 + \vec{p}^{\,2}} 
 \right] \,.
\end{eqnarray}
With the help of 
\begin{equation} 
\frac{1}{L^3} \sum_{\vec{p}} 1 - \int \! \frac{\mathrm d^3 p}{(2 \pi)^3} 1 = 0
\end{equation}
we obtain
\begin{eqnarray}
 & & \Delta_b (L) =
  \frac{3}{f_\pi^2} \int \! \frac{\mathrm d p_4}{2 \pi} 
 \left[ \frac{1}{L^3} \sum_{\vec{p}} 
  \frac{(2 c_1 - c_3) m_\pi^2 + c_2 p_4^2}
     { m_\pi^2 + p_4^2 + \vec{p}^{\,2}} 
\right. \nonumber \\ 
 & & \qquad \qquad \qquad \qquad {}  
  - \int \! \frac{\mathrm d^3 p}{(2 \pi)^3}  \left. 
\frac{(2 c_1 - c_3) m_\pi^2 + c_2 p_4^2}
     { m_\pi^2 + p_4^2 + \vec{p}^{\,2}} 
 \right] \,.
\nonumber  \\ & & 
\end{eqnarray}
Now we can use Eq.~(\ref{basic}) to rewrite this result as
\begin{eqnarray} 
 & & \frac{3}{f_\pi^2} \cdot \frac{1}{4 \pi^{3/2}} \sum_{\vec{n}} {}^\prime
\left( \frac{L^2 \vec{n}^{\,2}}{4} \right) ^{-1/4} 
\int \! \frac{\mathrm d p_4}{2 \pi} 
\frac{(2 c_1 - c_3) m_\pi^2 + c_2 p_4^2}
     {\left( m_\pi^2 + p_4^2 \right) ^{-1/4}}
\nonumber \\ & & \hspace*{6.0cm} {} \times
 K_{-1/2} \left( L |\vec{n}| \sqrt{m_\pi^2 + p_4^2} \right) \,,
\end{eqnarray}
and by means of Eq.~6.596.3 in~\cite{grad} we get the final formula
\begin{equation} \label{Deltab}
\Delta_b(L) = 
 \frac{3 m_\pi^4}{4 \pi^2 f_\pi^2} \sum_{\vec{n}} {}^\prime
\left[ (2 c_1 - c_3) \frac{K_1(L |\vec{n}| m_\pi)}{L |\vec{n}| m_\pi}
+ c_2 \frac{K_2(L |\vec{n}| m_\pi)}{(L |\vec{n}| m_\pi)^2} \right] \,.
\end{equation}
Hence we have
\begin{equation} \label{Delta}
m_N (L) - m_N (\infty) = \Delta_a(L) +  \Delta_b(L) + O(p^5) \,, 
\end{equation}
where $\Delta_a(L)$ is given in Eq.~(\ref{Deltaa}). Eq.~(\ref{Delta})
represents the complete $O(p^4)$ result for the volume dependence of the
nucleon mass in relativistic baryon chiral perturbation theory and will
serve as the basis for our numerical studies. Again, the finite volume
did not lead to the appearance of additional parameters. 

The leading contribution in Eq.~(\ref{Delta}) comes from the terms 
with $|\vec{n}|=1$ corresponding to pions which travel around the 
spatial box exactly once. In the Appendix we 
bring this contribution into the ``dispersive'' form introduced by
L\"uscher~\cite{luescher}, which relates the finite volume effects to 
the pion-proton forward elastic scattering amplitude. We find
\begin{eqnarray}
& & [m_N (L) - m_N (\infty)]\Big |_{\mathrm {leading}}
\nonumber \\ & & {}
 = \frac{9 g_A^2 m_0 m_\pi^2}{8 \pi^2 f_\pi^2} \Bigg\{
   \frac{\pi}{m_0 L} \exp \left( - m_\pi L 
                   \sqrt{1 - \frac{m_\pi^2}{4 m_0^2}} \right) 
\nonumber \\ & & {} \qquad \qquad \qquad 
 - \frac{1}{m_\pi L} \int_{- \infty}^\infty \! \mathrm d p \,
   \frac{\exp \left( - m_\pi L \sqrt{1+p^2} \right)}
        {1 + 4 m_0^2 \, p^2 / m_\pi^2}   \Bigg\}
\nonumber \\ & & {}
 + \frac{9m_\pi^4}{2 \pi^2 f_\pi^2} 
 \left[ (2 c_1 - c_3) \frac{K_1(m_\pi L)}{m_\pi L}
 + c_2 \frac{K_2(m_\pi L)}{(m_\pi L)^2} \right] + O(p^5) \,.
\end{eqnarray}
The coefficients $2 c_1 - c_3$ and $c_2$ are related to coefficients
in the so-called subthreshold expansion~\cite{hoehler}.
Unless $m_\pi L$ is rather large, the subleading terms with $|\vec{n}|>1$
are not negligible.  

\section{Monte Carlo data}

We want to compare our formulae to Monte Carlo data for nucleon 
masses. As the 
finite size effects rely essentially on the existence of a ``pion cloud''
and hence on the presence of sea quarks, such a comparison makes sense 
only for masses extracted from dynamical simulations. Large scale 
simulations with two flavours of dynamical quarks are beginning
to deliver results, but the quark masses are still rather large, and
the continuum limit is not easy to perform reliably, because only a
rather small range in the lattice spacing $a$ is covered. Therefore it is 
not straightforward to set the scale for the simulation results. A popular
method is to use the force parameter $r_0$~\cite{sommer} for that purpose, 
because it is expected to depend only weakly on the quark masses. In
the following we shall adopt this procedure using $r_0 = 0.5 \, \mbox{fm}$.
However, it must be noted that recently some doubts have been raised on 
the reliability of this method~\cite{cutoff}, and it would certainly 
be advantageous to avoid the scale problem completely by considering
dimensionless ratios, e.g., ratios of masses with the pion decay constant
and performing the whole analysis for these quantities. Another point
of concern is the question of whether the quark masses in the simulations
are small enough to justify the application of chiral perturbation theory.

Over the last years, the UKQCD and QCDSF collaborations have generated 
gauge field configurations with two flavours of dynamical quarks for 
a variety of quark masses, lattice spacings and volumes. These 
simulations use the standard Wilson plaquette action for the gauge 
fields and the non-perturbatively $O(a)$ improved clover action for 
the fermions. The algorithm employed is the Hybrid Monte Carlo 
algorithm, which recently could be sped up considerably~\cite{timur}. 
Details of the extraction of $r_0/a$ are given in~\cite{UKQCD}. 

The computation of the nucleon masses on the configurations generated
by UKQCD is also described in Ref.~\cite{UKQCD}. On the configurations
generated by QCDSF a somewhat different procedure has been used.
While standard interpolating fields have been employed in both cases, 
QCDSF has applied Jacobi smearing instead of fuzzing when computing the
quark propagators from point sources. The nucleon masses have been
obtained by fitting the source and sink smeared correlation function to
$
  A_N \exp\left[-m_N t\right] +
  A_{N^*} \exp\left[-m_{N^*} (T-t)\right] ,
$
where $N^*$ is the negative parity partner of the nucleon. 
The $\chi^2$ was calculated from the diagonal part of the covariance 
matrix only. But it has been checked that using the full covariance matrix 
gives consistent results. The fit range was fixed by looking for a region
where the results for $a m_N$ and $a m_{N^*}$ are independent
of the fit range.

\begin{table}[htb]
\caption{Simulation parameters and results from the UKQCD and QCDSF
collaborations.}
\label{tab:ukqcdsf}
\vspace{0.8cm}
\begin{center}
\begin{tabular}{rlllllll}
\hline
{} & \multicolumn{1}{c}{Coll.} & \multicolumn{1}{c}{$\beta $} 
& \multicolumn{1}{c}{$\kappa_{\mathrm {sea}}$} 
& \multicolumn{1}{c}{volume}
& \multicolumn{1}{c}{$r_0/a$} & \multicolumn{1}{c}{$a m_\pi$}
& \multicolumn{1}{c}{$a m_N$} \\
\hline
 1 & QCDSF   & 5.20 & 0.1342  & $16^3 \times 32$ & 4.077(31) & 0.5841(11)  
   & 1.1071(40) \\
 2 & UKQCD   & 5.20 & 0.1350  & $16^3 \times 32$ & 4.754(40) & 0.405(5)  
   & 0.883(12) \\
 3 & UKQCD   & 5.20 & 0.1355  & $16^3 \times 32$ & 5.041(40) & 0.294(4)  
   & 0.766(11)  \\
 4 & UKQCD   & 5.20 & 0.13565 & $16^3 \times 32$ & 5.246(51) & 0.2470(40)  
   & 0.676(24)  \\
 5 & UKQCD   & 5.20 & 0.1358  & $16^3 \times 32$ & 5.320(50) & 0.2080(70)  
   & 0.636(33)  \\
 6 & QCDSF   & 5.25 & 0.1346  & $16^3 \times 32$ & 4.737(21) & 0.4925(16)  
   & 0.9455(57) \\
 7 & UKQCD   & 5.25 & 0.1352  & $16^3 \times 32$ & 5.138(45) & 0.3842(16)  
   & 0.8032(69) \\
 8 & QCDSF   & 5.25 & 0.13575 & $24^3 \times 48$ & 5.430(60) & 0.2599(15)  
   & 0.6160(58) \\
 9 & UKQCD   & 5.26 & 0.1345  & $16^3 \times 32$ & 4.708(52) & 0.509(2)  
   & 1.011(10) \\
10 & UKQCD   & 5.29 & 0.1340  & $16^3 \times 32$ & 4.813(45) & 0.577(2)  
   & 1.086(9) \\
11 & QCDSF   & 5.29 & 0.1350  & $16^3 \times 32$ & 5.227(37) & 0.4208(8)   
   & 0.8344(34) \\
12 & QCDSF   & 5.29 & 0.1355  & $12^3 \times 32$ & 5.756(33) & 0.3637(48)  
   & 0.864(12)  \\
13 & QCDSF   & 5.29 & 0.1355  & $16^3 \times 32$ & 5.560(30) & 0.3334(15)  
   & 0.7188(76) \\
14 & QCDSF   & 5.29 & 0.1355  & $24^3 \times 48$ & 5.566(20) & 0.3265(6)   
   & 0.6857(35) \\
\hline
\end{tabular}
\end{center}
\vspace{0.8cm}
\end{table}

In addition we use data available in the literature~\cite{cppacs,jlqcd}. 
The CP-PACS collaboration~\cite{cppacs} works with a 
renormalisation-group improved gauge action and a mean field improved 
clover quark action, while the JLQCD collaboration~\cite{jlqcd} employs
the same actions as UKQCD and QCDSF. However, JLQCD uses a slightly
different value for the improvement parameter $c_{\mathrm {SW}}$. 
In Table~\ref{tab:ukqcdsf} we present our results including those that
have already been published in Ref.~\cite{UKQCD} (points 2, 3, 9, 10). 
Relevant data obtained by the CP-PACS and JLQCD 
collaborations are collected in Table~\ref{tab:japan}. Some important
quantities converted to physical units are given in Table~\ref{tab:phys}.

\begin{table}[htb]
\caption{Simulation parameters and results from the CP-PACS~\cite{cppacs}
and JLQCD~\cite{jlqcd} collaborations.}
\label{tab:japan}
\vspace{0.8cm}
\begin{center}
\begin{tabular}{rlllllll}
\hline
{} & \multicolumn{1}{c}{Coll.} & \multicolumn{1}{c}{$\beta $} 
& \multicolumn{1}{c}{$\kappa_{\mathrm {sea}}$} 
& \multicolumn{1}{c}{volume}
& \multicolumn{1}{c}{$r_0/a$} & \multicolumn{1}{c}{$a m_\pi$}
& \multicolumn{1}{c}{$a m_N$} \\
\hline
15 & CP-PACS & 1.95 & 0.1410 & $16^3 \times 32$ & 3.014(33) & 0.42700(98) 
   & 1.0532(51) \\
16 & CP-PACS & 1.95 & 0.1400 & $16^3 \times 32$ & 2.821(29) & 0.59580(69) 
   & 1.2679(39) \\
17 & CP-PACS & 1.95 & 0.1390 & $16^3 \times 32$ & 2.651(42) & 0.72857(68) 
   & 1.4559(38) \\
18 & CP-PACS & 1.95 & 0.1375 & $16^3 \times 32$ & 2.497(54) & 0.89400(52) 
   & 1.7035(34) \\
19 & CP-PACS & 2.10 & 0.1382 & $24^3 \times 48$ & 4.485(12) & 0.29459(85) 
   & 0.7204(42) \\
20 & CP-PACS & 2.10 & 0.1374 & $24^3 \times 48$ & 4.236(14) & 0.42401(46) 
   & 0.8955(35) \\
21 & CP-PACS & 2.10 & 0.1367 & $24^3 \times 48$ & 4.072(15) & 0.51671(67) 
   & 1.0226(32) \\
22 & CP-PACS & 2.10 & 0.1357 & $24^3 \times 48$ & 3.843(16) & 0.63010(61) 
   & 1.1855(26) \\
23 & CP-PACS & 2.20 & 0.1368 & $24^3 \times 48$ & 5.410(21) & 0.2785(22)  
   & 0.6314(55) \\
24 & CP-PACS & 2.20 & 0.1363 & $24^3 \times 48$ & 5.237(22) & 0.3554(10)  
   & 0.7349(42) \\
25 & CP-PACS & 2.20 & 0.1358 & $24^3 \times 48$ & 5.073(19) & 0.4190(13)  
   & 0.8252(47) \\
26 & CP-PACS & 2.20 & 0.1351 & $24^3 \times 48$ & 4.913(21) & 0.49996(83) 
   & 0.9330(76) \\
\hline
27 & JLQCD   & 5.20 & 0.1340 & $12^3 \times 48$ & 3.826(50) & 0.619(10)   
   & 1.153(12)  \\
28 & JLQCD   & 5.20 & 0.1343 & $12^3 \times 48$ & 4.031(84) & 0.5474(51)  
   & 1.094(23)  \\
29 & JLQCD   & 5.20 & 0.1346 & $12^3 \times 48$ & 4.200(56) & 0.5011(70)  
   & 1.006(24)  \\
30 & JLQCD   & 5.20 & 0.1350 & $12^3 \times 48$ & 4.481(67) & 0.4239(71)  
   & 0.915(34)  \\
31 & JLQCD   & 5.20 & 0.1355 & $12^3 \times 48$ & 5.06(12)  & 0.328(14)   
   & 0.820(36)  \\
32 & JLQCD   & 5.20 & 0.1340 & $16^3 \times 48$ & 3.880(36) & 0.6200(21)  
   & 1.166(20)  \\
33 & JLQCD   & 5.20 & 0.1343 & $16^3 \times 48$ & 4.098(45) & 0.5528(40)  
   & 1.063(13)  \\
34 & JLQCD   & 5.20 & 0.1346 & $16^3 \times 48$ & 4.287(58) & 0.4939(20)  
   & 0.991(14)  \\
35 & JLQCD   & 5.20 & 0.1350 & $16^3 \times 48$ & 4.621(42) & 0.4003(33)  
   & 0.828(12)  \\
36 & JLQCD   & 5.20 & 0.1355 & $16^3 \times 48$ & 5.059(71) & 0.2806(64)  
   & 0.707(29)  \\
37 & JLQCD   & 5.20 & 0.1340 & $20^3 \times 48$ & 3.946(30) & 0.61630(55) 
   & 1.1566(26) \\
38 & JLQCD   & 5.20 & 0.1343 & $20^3 \times 48$ & 4.143(29) & 0.55270(62) 
   & 1.0626(23) \\
39 & JLQCD   & 5.20 & 0.1346 & $20^3 \times 48$ & 4.336(50) & 0.49020(71) 
   & 0.9644(31) \\
40 & JLQCD   & 5.20 & 0.1350 & $20^3 \times 48$ & 4.635(53) & 0.40037(55) 
   & 0.8252(26) \\
41 & JLQCD   & 5.20 & 0.1355 & $20^3 \times 48$ & 5.092(83) & 0.27133(72) 
   & 0.6468(36) \\
\hline
\end{tabular}
\end{center}
\vspace{0.8cm}
\end{table}

\clearpage
\begin{table}[htb]
\caption{Simulation data converted to physical units using 
         $r_0 = 0.5 \mbox{fm}$.}
\label{tab:phys}
\vspace{0.8cm}
\begin{center}
\begin{tabular}{rccrll}
\hline
{} & $a$ [fm] & $L$ [fm] 
& \multicolumn{1}{c}{$m_\pi L$} & \multicolumn{1}{c}{$m_\pi$ [GeV]}
& \multicolumn{1}{c}{$m_N$ [GeV]} \\
\hline
 1 & 0.12 & 1.96 &  9.3 & 0.9398(74) & 1.781(15) \\
 2 & 0.11 & 1.68 &  6.5 & 0.760(11)  & 1.657(26) \\
 3 & 0.10 & 1.59 &  4.7 & 0.5849(92) & 1.524(25) \\
 4 & 0.10 & 1.52 &  4.0 & 0.5114(97) & 1.400(52) \\
 5 & 0.09 & 1.50 &  3.3 & 0.437(15)  & 1.335(70) \\
 6 & 0.11 & 1.69 &  7.9 & 0.9207(51) & 1.768(13) \\
 7 & 0.10 & 1.56 &  6.1 & 0.7791(76) & 1.629(20) \\
 8 & 0.09 & 2.21 &  6.2 & 0.5570(69) & 1.320(19) \\
 9 & 0.11 & 1.70 &  8.1 & 0.946(11)  & 1.878(28) \\
10 & 0.10 & 1.66 &  9.2 & 1.096(11)  & 2.063(26) \\
11 & 0.10 & 1.53 &  6.7 & 0.8681(64) & 1.721(14) \\
12 & 0.09 & 1.04 &  4.4 & 0.826(12)  & 1.963(29) \\
13 & 0.09 & 1.44 &  5.3 & 0.7316(51) & 1.577(19) \\
14 & 0.09 & 2.16 &  7.8 & 0.7172(29) & 1.5062(94) \\
\hline
15 & 0.17 & 2.65 &  6.8 & 0.5079(57) & 1.253(15) \\
16 & 0.18 & 2.84 &  9.5 & 0.6633(69) & 1.412(15) \\
17 & 0.19 & 3.02 & 11.7 & 0.762(12)  & 1.523(24) \\
18 & 0.20 & 3.20 & 14.3 & 0.881(19)  & 1.679(36) \\
19 & 0.11 & 2.68 &  7.1 & 0.5214(21) & 1.2751(82) \\
20 & 0.12 & 2.83 & 10.2 & 0.7088(25) & 1.4971(77) \\
21 & 0.12 & 2.95 & 12.4 & 0.8304(32) & 1.6434(79) \\
22 & 0.13 & 3.12 & 15.1 & 0.9556(41) & 1.7980(85) \\
23 & 0.09 & 2.22 &  6.7 & 0.5946(52) & 1.348(13) \\
24 & 0.10 & 2.29 &  8.5 & 0.7345(37) & 1.519(11) \\
25 & 0.10 & 2.37 & 10.1 & 0.8389(41) & 1.652(11) \\
26 & 0.10 & 2.44 & 12.0 & 0.9694(45) & 1.809(17) \\
\hline
27 & 0.13 & 1.57 &  7.4 & 0.935(19)  & 1.741(29) \\
28 & 0.12 & 1.49 &  6.6 & 0.871(20)  & 1.740(52) \\
29 & 0.12 & 1.43 &  6.0 & 0.831(16)  & 1.667(46) \\
30 & 0.11 & 1.34 &  5.1 & 0.750(17)  & 1.618(65) \\
31 & 0.10 & 1.19 &  3.9 & 0.655(32)  & 1.637(82) \\
32 & 0.13 & 2.06 &  9.9 & 0.9494(94) & 1.785(35) \\
33 & 0.12 & 1.95 &  8.8 & 0.894(12)  & 1.719(28) \\
34 & 0.12 & 1.87 &  7.9 & 0.836(12)  & 1.677(33) \\
35 & 0.11 & 1.73 &  6.4 & 0.7300(90) & 1.510(26) \\
36 & 0.10 & 1.58 &  4.5 & 0.560(15)  & 1.412(61) \\
37 & 0.13 & 2.53 & 12.3 & 0.9598(74) & 1.801(14) \\
38 & 0.12 & 2.41 & 11.1 & 0.9037(64) & 1.737(13) \\
39 & 0.12 & 2.31 &  9.8 & 0.8388(98) & 1.650(20) \\
40 & 0.11 & 2.16 &  8.0 & 0.7324(84) & 1.509(18) \\
41 & 0.10 & 1.96 &  5.4 & 0.5453(90) & 1.300(22) \\
\hline
\end{tabular}
\end{center}
\vspace{0.8cm}
\end{table}

\clearpage

\section{Comparison with chiral perturbation theory}

In this section we confront the nucleon masses presented in the 
preceding section with chiral perturbation theory. We use data obtained 
on (relatively) large lattices to find appropriate values for
the parameters of the chiral expansion. After fixing these parameters,
the finite size formulae predict the volume dependence of the nucleon
masses without any free parameter. 

In Ref.~\cite{nmass} an analysis of data points from 
Tables~\ref{tab:ukqcdsf} and \ref{tab:japan} obtained on 
(relatively) large and fine lattices has been presented 
(see also Ref.~\cite{meissner}). 
More precisely, masses from simulations
with $a < 0.15 \, \mbox{fm}$, $m_\pi L > 5$ and $m_\pi < 800 \, \mbox{MeV}$
have been considered. Here we shall impose the same conditions and
select the data points 2, 7, 8, 14, 19, 20, 23, 24, 40, 41 for the
further analysis. (Point 13, which also fulfills the above criteria, 
is discarded, because the corresponding simulation parameters agree 
with those of point 14, except that the lattice is smaller. Similarly,
simulations 30 and 35 just repeat simulation 40 on smaller lattices.) 

We cannot determine all parameters in the chiral 
perturbation theory formulae from fits to the lattice data. Therefore 
we decided to fix $g_A$, $f_\pi$, $c_2$, $c_3$, and to fit the remaining
parameters $m_0$, $c_1$, $e_1^r(\lambda)$ choosing 
$\lambda = 1 \, \mbox{GeV}$.  
For $g_A$ and $f_\pi$ we take the physical values $g_A = 1.267$,
$f_\pi = 92.4 \, \mbox{MeV}$. Choosing values for $c_2$ and $c_3$
is more difficult, although there are quite a few phenomenological 
determinations in the literature (see, e.g., 
\cite{bernard,fettes2,fettes,mojzis,butt,beleu2,rent,entem,epel}). 
In particular, $c_3$ seems to be subject to a considerable uncertainty. 
We set $c_2 = 3.2 \, \mbox{GeV}^{-1}$, which is 
compatible with Refs.~\cite{bernard,fettes2,fettes,mojzis,entem}, 
and consider two possibilities for $c_3$: $c_3 = -3.4 \, \mbox{GeV}^{-1}$ 
and $c_3 = -4.7 \, \mbox{GeV}^{-1}$. According to Refs.~\cite{entem,epel}, 
the value $c_3 = -3.4 \, \mbox{GeV}^{-1}$ is consistent with the empirical 
nucleon-nucleon phase shifts and with the value extracted in \cite{butt} 
from pion-nucleon scattering. On the other hand, 
$c_3 = -4.7 \, \mbox{GeV}^{-1}$ is the central value obtained in the 
pion-nucleon scattering analysis of Ref.~\cite{butt}, albeit with large
error bars, and is consistent with the results of Ref.~\cite{rent}.

\begin{figure}[htbp]
\begin{center}
\epsfig{file=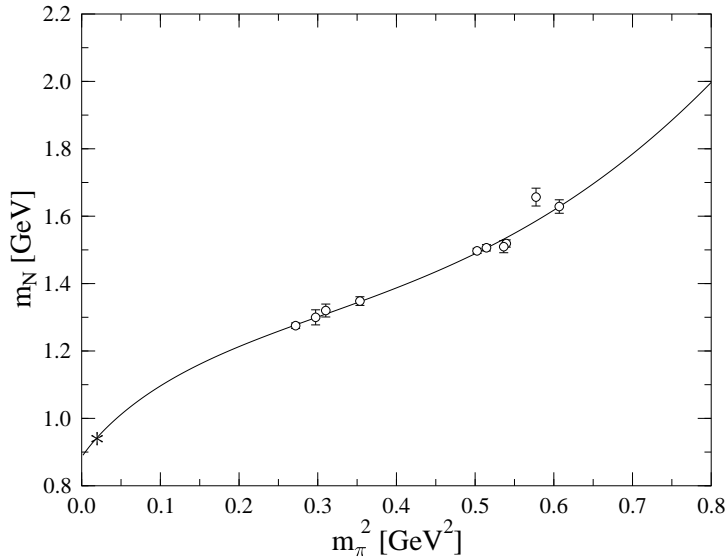,width=12cm}
\caption{Nucleon mass data on (relatively) large and fine lattices.
         The star indicates the physical point. The curve corresponds
         to Fit 1 in Table~\ref{tab:fitresults}.}
\label{fig:mnfit}
\end{center}
\end{figure}

From a fit of Eq.~(\ref{mnp4}) to the ten points selected above with 
$c_3 = -3.4 \, \mbox{GeV}^{-1}$ we obtain the results
labelled as ``Fit 1'' in Table~\ref{tab:fitresults}. Data points
and fit are shown in Fig.~\ref{fig:mnfit}. (Using $g_A = 1.2$, 
$f_\pi = 88 \, \mbox{MeV}$ as better approximations to the values 
in the chiral limit does not lead to significant differences.)
It is first of all remarkable 
that the Monte Carlo data obtained by different collaborations with 
different actions all lie close to a single curve. 
Thus there seems to be no sign of appreciable lattice artefacts.
Secondly, the fit describes the data very well and is at the same time 
compatible with the physical mass values. Thirdly, the fit parameters
take values which are consistent with known phenomenology. 
In particular, $c_1$ compares favourably with phenomenological 
determinations~\cite{bernard,mojzis,butt,beleu2}.
Repeating the fit with $c_3 = -4.7 \, \mbox{GeV}^{-1}$ leads to
the results labelled as ``Fit 2'' in Table~\ref{tab:fitresults},
which are farther away from the phenomenological numbers. 

\begin{table}[htb]
\caption{Results from the four fits of nucleon mass data as 
         described in the text. The parameter $c_3$ has been kept fixed.}
\label{tab:fitresults}
\vspace{0.8cm}
\begin{center}
\begin{tabular}{cccccr@{.}l}
\hline
{} & $c_3$ [GeV$^{-1}$] & $m_0$ [GeV] & $c_1$ [GeV$^{-1}$] 
& $e_1^r(\lambda = 1 \, \mbox{GeV})$ [GeV$^{-3}$] 
& \multicolumn{2}{c}{$\chi^2$} \\
\hline
Fit 1 & $-3.4$ & 0.89(6) & $-0.93(5)$ & 2.8(4) & 12 & 18 \\
Fit 2 & $-4.7$ & 0.76(6) & $-1.25(5)$ & 1.7(5) & 11 & 85 \\
Fit 3 & $-3.4$ & 0.88(-) & $-0.93(4)$ & 3.0(6) & 0  & 29 \\
Fit 4 & $-4.7$ & 0.87(-) & $-1.11(4)$ & 3.2(6) & 0  & 39 \\
\hline
\end{tabular}
\end{center}
\vspace{0.8cm}
\end{table}

Of course, one may doubt the applicability of chiral perturbation 
theory at the larger pion masses which enter our analysis. On the other
hand, we do not have enough results at smaller masses to restrict 
ourselves to a safer mass range performing the same kind of fits as 
above. We can, however, constrain the fit by requiring the fit curve 
to pass through the physical point. We implement
this constraint by choosing $m_0$ such that the condition is satisfied.
In this way we eliminate one fit parameter leaving only $c_1$ and 
$e_1^r(\lambda = 1 \, \mbox{GeV})$ to be fitted. The reduced number of
parameters then allows us to restrict the fit to a smaller number of masses.
From such constrained fits to the four data points at the smallest masses 
(points 8, 19, 23, 41) we get the results labelled as ``Fit 3'' 
(with $c_3 = -3.4 \, \mbox{GeV}^{-1}$) and ``Fit 4''
(with $c_3 = -4.7 \, \mbox{GeV}^{-1}$) in Table~\ref{tab:fitresults}.
While the results of Fit 1 and Fit 3 are well compatible with each other
(and with phenomenology), the constraint has a stronger effect for 
$c_3 = -4.7 \, \mbox{GeV}^{-1}$. 

\begin{figure}[htbp]
\begin{center}
\epsfig{file=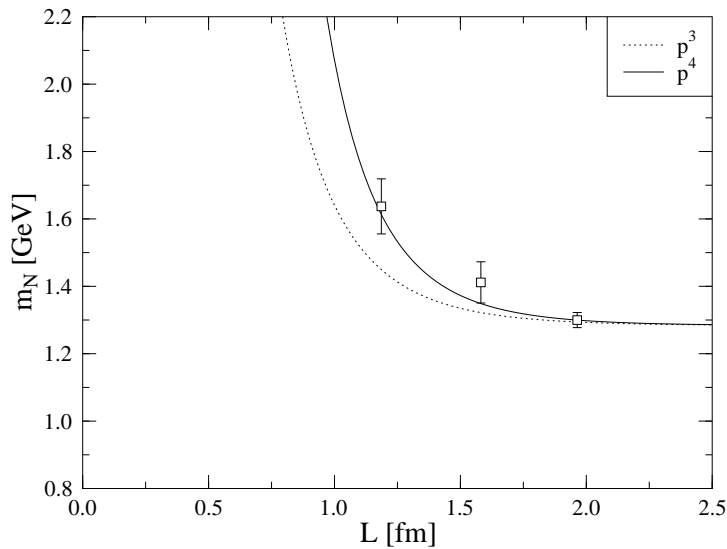,width=12cm}
\caption{Volume dependence of the nucleon mass for 
         $m_\pi = 545 \, \mbox{MeV}$ (data points 31, 36, 41). The dotted 
         curve shows the contribution of the $p^3$ term, while the 
	 solid curve includes also the $p^4$ correction, with the 
         parameters taken from Fit 1.}
\label{fig:mpi545fv}
\end{center}
\end{figure}

\begin{figure}[htbp]
\begin{center}
\epsfig{file=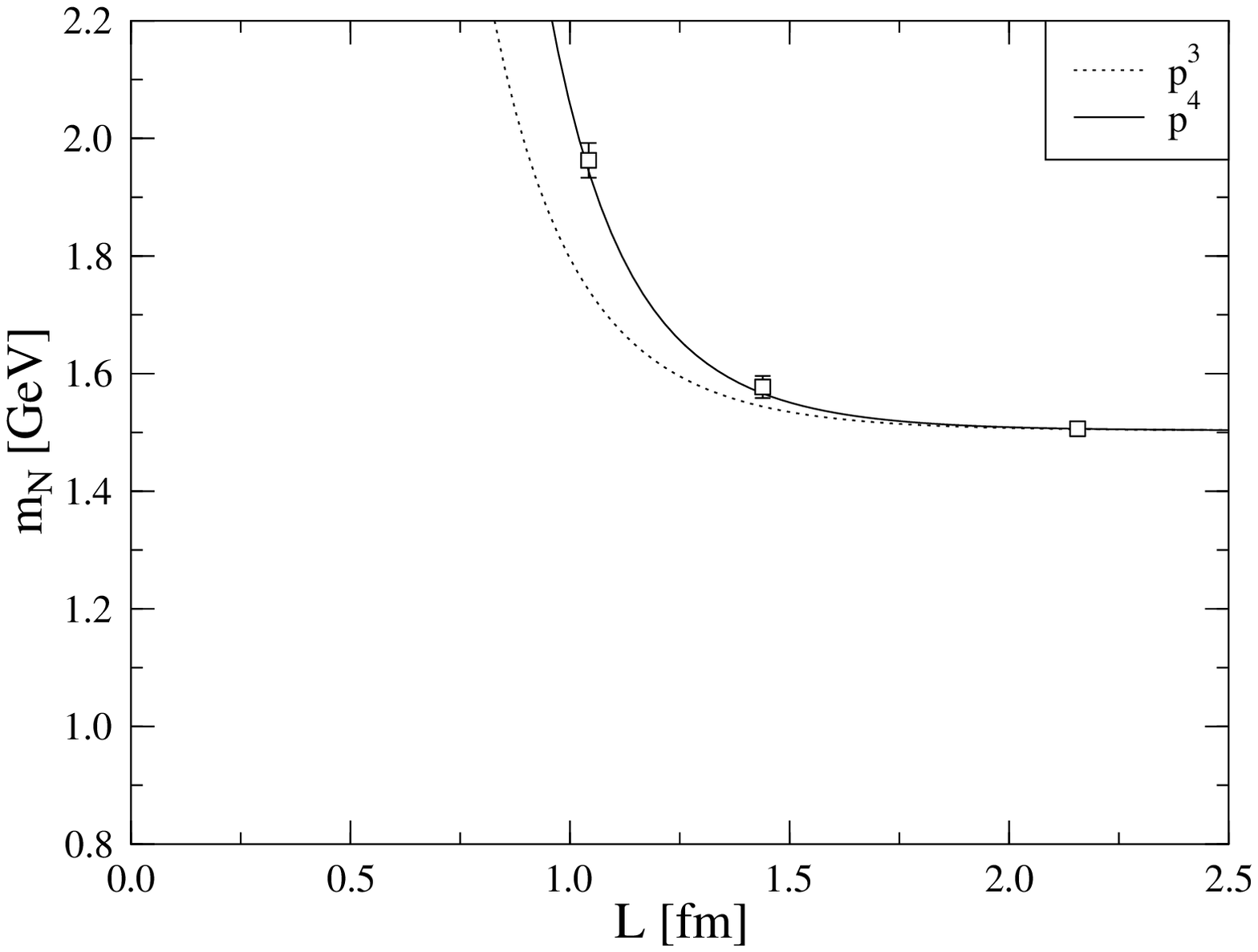,width=12cm}
\caption{Volume dependence of the nucleon mass for 
         $m_\pi = 717 \, \mbox{MeV}$ (data points 12, 13, 14). The dotted 
         curve shows the contribution of the $p^3$ term, while the 
	 solid curve includes also the $p^4$ correction, with the 
         parameters taken from Fit 1.}
\label{fig:mpi717fv}
\end{center}
\end{figure}

\begin{figure}[htbp]
\begin{center}
\epsfig{file=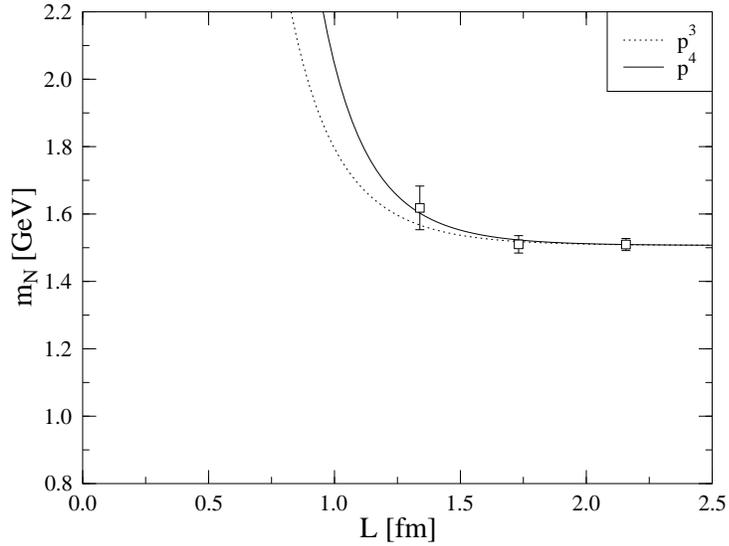,width=12cm}
\caption{Volume dependence of the nucleon mass for 
         $m_\pi = 732 \, \mbox{MeV}$ (data points 30, 35, 40). The dotted 
         curve shows the contribution of the $p^3$ term, while the 
	 solid curve includes also the $p^4$ correction, with the 
         parameters taken from Fit 1.}
\label{fig:mpi732fv}
\end{center}
\end{figure}

Using the parameters obtained from Fit 1, which are 
consistent with phenomenology and the large volume Monte Carlo data, 
we can now evaluate our finite size corrections. Note that no further 
free parameters are involved. For the comparison with the Monte Carlo 
results we need data from simulations which differ only in the lattice 
size while all other parameters are kept fixed. There are six sets of 
three simulations each among the results in Tables~\ref{tab:ukqcdsf}, 
\ref{tab:japan}. We choose the three sets with the smallest pion 
masses and plot the nucleon masses versus the lattice size in 
Figs.~\ref{fig:mpi545fv}, \ref{fig:mpi717fv}, \ref{fig:mpi732fv}. 
The curves are computed from our finite size formulae. The curve 
labelled $p^4$ corresponds to 
\begin{equation}
m_N(L) = \Delta_a (L) + \Delta_b (L) + m_N(\infty) \,,
\end{equation}
where $m_N(\infty)$ is determined such that $m_N(L)$ on the largest lattice 
agrees with the Monte Carlo value. For the pion mass we use the value
from the largest lattice. In the curve labelled $p^3$, the $p^4$
contribution $\Delta_b (L)$ has been left out, but $m_N(\infty)$ has 
not been changed.

Our formula describes the volume dependence of the nucleon mass 
remarkably well. However, the agreement would deteriorate
had we taken into account only pions which travel around the lattice 
exactly once, by restricting the sums over $\vec{n}$ in 
Eqs.~(\ref{Deltaa}), (\ref{Deltab}) to $\vec{n}$ with $|\vec{n}|=1$.

\section{Conclusions}

Chiral effective field theories evaluated in a finite volume yield expressions
for physical quantities, e.g.\ masses, in the form of an expansion with
an expansion parameter $p$, where the pion mass $m_\pi$ and the inverse 
box size $1/L$ both count as quantities of order $p$. We have applied
this formalism to the nucleon mass using relativistic SU(2)$_\mathrm f$
baryon chiral perturbation theory up to and including $O(p^4)$ terms.

Setting the scale with $r_0 = 0.5 \, \mbox{fm}$, we can convert Monte Carlo
results obtained in simulations with dynamical quarks to physical units
and compare them with our formulae. The infinite-volume formulae lead
to a surprisingly good description of the data coming from (relatively) 
large lattices, with values for the coupling constants of the effective 
theory which are completely consistent with known 
phenomenology~\cite{nmass,meissner}. Given these coupling constants, there are 
no free parameters left in the finite size formulae. But the $p^4$ 
formulae reproduce the finite size effects observed in the Monte Carlo 
data surprisingly well. For this agreement it is essential that our 
formalism incorporates the effects of pions travelling around the 
lattice arbitrarily many times. Thus we arrive at a more optimistic
opinion on the applicability of chiral perturbation theory than previous
investigations~\cite{fukugita}.

We find a remarkably consistent picture of the volume and pion mass 
dependence of the nucleon mass, in spite of the fact that the 
``convergence'' of the chiral expansion at the rather large pion masses
used in the available simulations is far from obvious. Lattice data
for smaller masses will therefore be of great importance to corroborate
this conclusion. Another point which deserves further investigation 
is the cutoff and mass dependence of $r_0$~\cite{cutoff}. Alternatively 
one could try to avoid the use of $r_0$ by basing the analysis on 
ratios of masses with a quantity which is more easily accessible to 
chiral perturbation theory than $r_0$, e.g.\ the pion decay constant. 
One could then work out formulae directly for these ratios. Finally 
it would be interesting to see if the pion mass can be described 
along similar lines.

\section*{Appendix}
In this Appendix we compare our description of finite size effects with
a finite size formula derived by L\"uscher~\cite{luescher}. He expresses
the leading contributions to the relative finite size effect
\begin{equation}
\delta_N = \left(m_N (L) - m_N (\infty) \right)/ m_N (\infty)
\end{equation}
in terms of the pion-nucleon coupling constant $g_{\pi N}$ with the 
physical value
\begin{equation}
\frac{g_{\pi N}^2}{4 \pi} \approx 14.3 
\end{equation}
and the pion-proton forward elastic scattering amplitude $F_{\pi p}(\nu)$. 
Here $\nu$ is the ``crossing variable'' 
$\nu = (s-u)/(4 m_N)$ with $m_N = m_N (\infty)$. L\"uscher finds 
\begin{eqnarray} 
& & \delta_N^{\mbox{\scriptsize L\"uscher}} = 
\frac{9}{16 \pi m_N L} \left( \frac{m_\pi}{m_N} \right)^2 g_{\pi N}^2 
 \exp \left( - m_\pi L \sqrt{1 - \frac{m_\pi^2}{4 m_N^2}} \right) 
\nonumber \\ & &  \quad {}
  - \frac{3}{8 \pi m_\pi L} \left( \frac{m_\pi}{m_N} \right)^2
   \int_{- \infty}^\infty \! \frac{\mathrm d p}{2 \pi} \,
   \exp \left( - m_\pi L \sqrt{1+p^2} \right) F_{\pi p}(\mathrm i m_\pi p)
\nonumber \\[0.3cm] & &  \quad {}
   + O \left( \mathrm e ^{- \alpha m_\pi L} \right) \,,
\label{luescher}
\end{eqnarray}
where $\alpha \geq \sqrt{3/2}$. The amplitude $F_{\pi p}(\nu)$ is related
to the amplitude $C^+(\nu) = D^+(\nu)$ of Ref.~\cite{hoehler} through
\begin{equation}
F_{\pi p}(\nu) = 6 m_N C^+(\nu) \,.
\end{equation}
It can conveniently be decomposed into the pseudovector Born term and 
a remainder $R(\nu)$:
\begin{equation}
F_{\pi p}(\nu) = \frac{6 g_{\pi N}^2}{1 - 4 m_N^2 \nu^2/m_\pi^4} + R(\nu) \,.
\end{equation}
Using the so-called subthreshold expansion~\cite{hoehler} for vanishing
momentum transfer
\begin{equation}
R(\nu) = 6 m_N \sum_{k=0}^\infty d_{k0}^+ \nu^{2k}
\end{equation}
L\"uscher's formula becomes (ignoring questions of convergence)
\begin{eqnarray} 
& & \delta_N^{\mbox{\scriptsize L\"uscher}} = 
\frac{9}{8 \pi^2} \left( \frac{m_\pi}{m_N} \right)^2 g_{\pi N}^2 
\Bigg\{ \frac{\pi}{2 m_N L} \exp \left( - m_\pi L 
                   \sqrt{1 - \frac{m_\pi^2}{4 m_N^2}} \right) 
\nonumber \\ & & \qquad \qquad {}
  -   \frac{1}{m_\pi L} \int_{- \infty}^\infty \! \mathrm d p \,
   \frac{\exp \left( - m_\pi L \sqrt{1+p^2} \right)}
        {1 + 4 m_N^2 \, p^2 / m_\pi^2} \Bigg \} 
\nonumber \\ & & \qquad \qquad {}
 -\frac{9}{4 \pi} \frac{m_\pi}{m_N L} \sum_{k=0}^\infty m_\pi^{2k} d_{k0}^+
  (-1)^k \int_{- \infty}^\infty \! \frac{\mathrm d p}{2 \pi} \,
  \exp \left( - m_\pi L \sqrt{1+p^2} \right) p^{2k}
\nonumber \\[0.3cm] & & \qquad \qquad {}
   + O \left( \mathrm e ^{- \alpha m_\pi L} \right) \,.
\label{deltaN}
\end{eqnarray}

As it is written, L\"uscher's formula refers to a fixed value of
$m_\pi$. However, as far as we have information on the quark-mass
dependence of $m_N$, $g_{\pi N}$ and the scattering amplitude --
and chiral perturbation theory provides such information -- we can 
make use of this knowledge to describe the $m_\pi$ dependence of
$\delta_N$ as well. But L\"uscher's formula takes into account 
only pions which travel around the lattice exactly once. For the 
comparison with our formulae we therefore have 
to restrict the sums over $\vec{n}$ in Eqs.~(\ref{Deltaa}), (\ref{Deltab})
to $\vec{n}$ with $|\vec{n}|=1$. In order to get the relative finite 
size effect we must furthermore divide by $m_N$. Because our finite 
size formulae correspond to $O(p^3)$ (in the case of $\Delta_a$) and 
$O(p^4)$ (if also $\Delta_b$ is taken into account) in the chiral 
expansion, we identify here and in the following $m_N$ with $m_0$ 
-- the higher terms in $m_N$ would give a contribution only at $O(p^5)$.
Thus we get from graph (a)
\begin{equation} \label{deltaa}
\delta_a = \frac{3 g_A^2 m_\pi^2}{16 \pi^2 f_\pi^2}
  \int_0^\infty \! \mathrm d x \, \cdot 6 \cdot \, 
  K_0 \left( L \sqrt{m_0^2 x^2 + m_\pi^2 (1-x)} \right) \,,
\end{equation}
while graph (b) yields
\begin{equation} \label{deltab}
\delta_b = \frac{9}{2 \pi^2 f_\pi^2} \frac{m_\pi^4}{m_0}
\left[ (2 c_1 - c_3) \frac{K_1(m_\pi L)}{m_\pi L}
+ c_2 \frac{K_2(m_\pi L)}{(m_\pi L)^2} \right] \,.
\end{equation}

First we consider $\delta_a$ and show that the integral over $x$ 
in (\ref{deltaa}) can be rewritten as
\begin{eqnarray} 
& & \int_0^\infty \! \mathrm d x \, 
     K_0 \left( L \sqrt{m_0^2 x^2 + m_\pi^2 (1-x)} \right)
\nonumber \\ & & {}
 = \frac{\pi}{m_0 L} \exp \left( - m_\pi L 
                   \sqrt{1 - \frac{m_\pi^2}{4 m_0^2}} \right)
 - \frac{1}{m_\pi L} \int_{- \infty}^\infty \! \mathrm d p \,
   \frac{\exp \left( - m_\pi L \sqrt{1+p^2} \right)}
        {1 + 4 m_0^2 \, p^2 / m_\pi^2} 
\nonumber \\ & & 
\label{paul}
\end{eqnarray}
for $m_\pi^2 < 4 m_0^2$. We begin by writing
\begin{eqnarray} 
& & \int_0^\infty \! \mathrm d x \, 
     K_0 \left( L \sqrt{m_0^2 x^2 + m_\pi^2 (1-x)} \right)
\nonumber \\ & & {}
= \int_{- \infty}^\infty \! \mathrm d x \, 
     K_0 \left( L \sqrt{m_0^2 x^2 + m_\pi^2 (1-x)} \right)
\nonumber \\ & & \hspace*{4.5cm} {}
 -  \int_{- \infty}^0 \! \mathrm d x \, 
     K_0 \left( L \sqrt{m_0^2 x^2 + m_\pi^2 (1-x)} \right) \,.
\nonumber \\ & & {}
\end{eqnarray}
The first integral on the right-hand side can be evaluated with the 
help of formula~6.596.3 
in~\cite{grad}: 
\begin{eqnarray} 
& & \int_{- \infty}^\infty \! \mathrm d x \, 
     K_0 \left( L \sqrt{m_0^2 x^2 + m_\pi^2 (1-x)} \right)
\nonumber \\ & & \quad {}
= \int_{- \infty}^\infty \! \mathrm d x \, 
     K_0 \left( L \sqrt{m_0^2 x^2 + 
             m_\pi^2 \left( 1 -  \frac{m_\pi^2}{4 m_0^2} \right) } \right)
\nonumber \\ & & \quad {}
 = \frac{\pi}{m_0 L} \exp \left( - m_\pi L 
                   \sqrt{1 - \frac{m_\pi^2}{4 m_0^2}} \right) \,.
\end{eqnarray}
In the second integral we use the integral representation
\begin{equation}
K_0 (xy) = \int_y^\infty \! \mathrm d s \, \left( s^2 - y^2 \right) ^{-1/2}
\mathrm e ^{-xs}
\end{equation}
(valid for $x,y > 0$) and interchange the order of the integrations:
\begin{eqnarray} 
& & \int_{- \infty}^0 \! \mathrm d x \, 
     K_0 \left( L \sqrt{m_0^2 x^2 + m_\pi^2 (1-x)} \right)
\nonumber \\ & & {}
= \int_1^\infty \! \mathrm d s \int_0^{y(s)} \! \mathrm d x
\left( s^2 - 1 - \frac{m_0^2}{m_\pi^2} x^2 - x \right) ^{-1/2}
\mathrm e ^{- m_\pi L s} 
\end{eqnarray}
with the abbreviation 
\begin{equation}
y(s) = \frac{m_\pi}{m_0} \sqrt{s^2 - 1 + \frac{m_\pi^2}{4 m_0^2}}
  - \frac{m_\pi^2}{2 m_0^2} \,.
\end{equation}
The integration over $x$ can now be performed leading to
\begin{equation}
\frac{m_\pi}{m_0} \int_1^\infty \! \mathrm d s \, \mathrm e ^{- m_\pi L s}
\arctan \left( \frac{2 m_0}{m_\pi} \sqrt{s^2-1} \right) \,.
\end{equation}
Partial integration yields
\begin{equation}
\frac{2}{m_\pi L} \int_1^\infty \! \mathrm d s \, 
\frac{\mathrm e ^{- m_\pi L s}}{1 + 4m_0^2 (s^2-1)/m_\pi^2}
\cdot \frac{s}{\sqrt{s^2-1}} \,,
\end{equation}
and the change of variables $s=\sqrt{1+p^2}$ gives the final expression
\begin{equation} 
\int_{- \infty}^0 \! \mathrm d x \, 
     K_0 \left( L \sqrt{m_0^2 x^2 + m_\pi^2 (1-x)} \right)
 =   \frac{1}{m_\pi L} \int_{- \infty}^\infty \! \mathrm d p \,
   \frac{\exp \left( - m_\pi L \sqrt{1+p^2} \right)}
        {1 + 4 m_0^2 \, p^2 / m_\pi^2} 
\end{equation}
completing the proof of Eq.~(\ref{paul}).

With the help of the Goldberger-Treiman relation
\begin{equation} \label{goldtr}
g_A =  g_{\pi N} \frac{f_\pi}{m_0} \,,
\end{equation}
which is exact in the chiral limit, we now obtain
\begin{eqnarray} 
& & \delta_a = 
\frac{9}{8 \pi^2} \left( \frac{m_\pi}{m_0} \right)^2 g_{\pi N}^2 
\Bigg\{ \frac{\pi}{m_0 L} \exp \left( - m_\pi L 
                   \sqrt{1 - \frac{m_\pi^2}{4 m_0^2}} \right) 
\nonumber \\ & & \hspace*{4.0cm} {}
  -   \frac{1}{m_\pi L} \int_{- \infty}^\infty \! \mathrm d p \,
   \frac{\exp \left( - m_\pi L \sqrt{1+p^2} \right)}
        {1 + 4 m_0^2 \, p^2 / m_\pi^2} \Bigg \} \,.
\end{eqnarray}
The chiral corrections to (\ref{goldtr}) are of order $p^2$ and hence
would contribute only at $O(p^5)$, beyond the accuracy of our calculation.
Thus $\delta_a$ reproduces the first two terms in (\ref{deltaN}) up to 
a factor of two in the first term, which in L\"uscher's treatment 
originates from nucleon exchange diagrams.

Do the following terms of Eq.~(\ref{deltaN}) contain our $\delta_b$, 
Eq.~(\ref{deltab})? With the help of 
\begin{equation} 
\int_{- \infty}^\infty \! \frac{\mathrm d p}{2 \pi} \,
  \exp \left( - z \sqrt{1+p^2} \right) p^{2k}
  = - \frac{1}{\pi} \frac{\Gamma (k+1/2)}{\Gamma (1/2)} 2^k
  \frac{\mathrm d}{\mathrm d z} z^{-k} K_k (z)
\end{equation}
and the tree-level relations (see, e.g., Ref.~\cite{beleu})
\begin{eqnarray} 
d_{00}^+ & = & - \frac{2 m_\pi^2}{f_\pi^2} (2 c_1 - c_3) + O(m_\pi^3) \,,
\nonumber \\
d_{10}^+ & = & \frac{2}{f_\pi^2} c_2 + O(m_\pi) 
\end{eqnarray}
we obtain 
\begin{eqnarray} 
& &  -\frac{9}{4 \pi} \frac{m_\pi}{m_0 L} \sum_{k=0}^\infty m_\pi^{2k} d_{k0}^+
  (-1)^k \int_{- \infty}^\infty \! \frac{\mathrm d p}{2 \pi} \,
  \exp \left( - m_\pi L \sqrt{1+p^2} \right) p^{2k}
\nonumber \\ & & \qquad \qquad {}
 = \frac{9}{2 \pi^2 f_\pi^2} \frac{m_\pi^4}{m_0}
 \left[ (2 c_1 - c_3) \frac{K_1(m_\pi L)}{m_\pi L}
 + c_2 \frac{K_2(m_\pi L)}{(m_\pi L)^2} + \cdots \right] 
\end{eqnarray}
in complete agreement with (\ref{deltab}). 

In short, when we write the $|\vec{n}|=1$ contributions of our 
finite size formula (\ref{Delta}) in the form that L\"uscher 
uses~\cite{luescher}, we find, within the accuracy of our chiral expansion,
\begin{eqnarray} 
& & \delta_N = 
\frac{9}{8 \pi m_N L} \left( \frac{m_\pi}{m_N} \right)^2 g_{\pi N}^2 
 \exp \left( - m_\pi L \sqrt{1 - \frac{m_\pi^2}{4 m_N^2}} \right) 
\nonumber \\ & &  \quad {}
  - \frac{3}{8 \pi m_\pi L} \left( \frac{m_\pi}{m_N} \right)^2
   \int_{- \infty}^\infty \! \frac{\mathrm d p}{2 \pi} \,
   \exp \left( - m_\pi L \sqrt{1+p^2} \right) F_{\pi p}(\mathrm i m_\pi p) \,,
\end{eqnarray}
which looks like L\"uscher's formula, except that the leading term has a
coefficient twice as large. Numerically this makes a significant difference
(more than a factor 2) because in L\"uscher's case the exponential and
the integral almost cancel each other.

\section*{Acknowledgements}

The numerical calculations have been performed on the Hitachi SR8000 at
LRZ (Munich), on the Cray T3E at EPCC (Edinburgh) under 
PPARC grant PPA/G/S/1998/00777, on the Cray T3E at NIC (J\"ulich) and ZIB
(Berlin), as well as on the APE1000 and Quadrics at DESY (Zeuthen). We
thank all institutions for their support.

This work has been supported in part by 
the European Community's Human Potential Program under contract 
HPRN-CT-2000-00145, Hadrons/Lattice QCD and
by the DFG (Forschergruppe Gitter-Hadronen-Ph\"anomenologie). 
A. A.K. thanks the DFG for a research grant (No.\ AL 596/1). 
We are also thankful for the support provided by the ECT$^*$ (Trento).
Discussions with A. Sch\"afer and W. Weise are gratefully acknowledged.
TRH thanks the Institute for Theoretical Physics of the University 
of Regensburg, the Institute for Physics of the Humboldt University
Berlin and DESY Zeuthen for their kind hospitality.

\end{document}